# Proximity-induced superconducting gap in the intrinsic magnetic topological insulator MnBi$_2$Te$_4$

Wen-Zheng Xu,[1] Chun-Guang Chu,[1] Zhen-Cun Pan,[1] Jing-Jing Chen,[2,3] An-Qi Wang,[1] Zhen-Bing Tan,[2,3,*]
Peng-Fei Zhu,[1] Xing-Guo Ye,[1] Da-Peng Yu,[2,3] and Zhi-Min Liao[1,†]

[1]*State Key Laboratory for Mesoscopic Physics and Frontiers Science Center for Nano-optoelectronics, School of Physics,
Peking University, Beijing 100871, China*
[2]*Shenzhen Institute for Quantum Science and Engineering, Southern University of Science and Technology, Shenzhen 518055, China*
[3]*International Quantum Academy, Shenzhen 518048, China*

We report magnetotransport measurements in the NbN-magnetic topological insulator MnBi$_2$Te$_4$ (MBT)-NbN junction at low temperature. At 10 mK, the nonlinear current-voltage characteristic of the junction shows a tunneling behavior, indicating the existence of interfacial potential barriers within the heterostructure. Under an out-of-plane perpendicular magnetic field, a transition from negative to positive magnetoresistance (MR) is found when increasing the bias voltage. A proximity-induced superconducting gap $\Delta_{ind}$ is estimated to be $\sim$0.1 meV by a pair of differential resistance dips. Moreover, the induced gap is enhanced by gradually tuning the Fermi level toward the charge neutral point by a back-gate voltage, which is ascribed to the increased transport contribution of the topological surface states in MBT. Intriguingly, the induced gap exhibits an anomalous magnetic-field-assisted enhancement, which may originate from the spin-orbit coupling and magnetic order of MBT. Our results reveal the interplay between magnetism and superconductivity in MBT, paving the way for further studies on topological superconductivity and chiral Majorana edge modes in quantum anomalous Hall insulator–superconductor hybrid systems.

## I. INTRODUCTION

Artificial superconductor heterostructure has aroused great research enthusiasm in condensed matter physics recently for its abundant exotic quantum phenomena [1–5]. The transport in the low-bias voltage regime at a normal metal (N)-superconductor (S) interface is governed by a quantum mechanism called Andreev reflection (AR). An incoming electron is retroreflected as a hole of opposite spin, and a spin-singlet Cooper pair is injected into the superconductor, thus doubling the electrical conductance [6]. In a S-N-S Josephson junction, the formation of Andreev bound states contributes to the presence of supercurrent. Transport properties may change drastically, however, when a ferromagnetic (FM) layer is introduced, considering its spin-splitting bands caused by the exchange interaction. It is predicted that the FM order suppresses Andreev retroreflection but enhances specular AR [7]. Additionally, signatures of spin-triplet pairing have been observed in S-FM-S junctions [8–13]. Moreover, the hybrid structure of a quantum anomalous Hall insulator (QAHI) and an *s*-wave superconductor has been predicted to host chiral Majorana edge modes (CMEMs) [14–16], which is a potential candidate material system to realize topological quantum computation. However, the finite net magnetization of FM hinders the application in nanoscale superconducting devices. Antiferromagnetic (AFM) compounds may be an appropriate substitute due to their negligible stray fields and unique properties in spintronics [17–21] and topological phases [22,23].

Recently, MnBi$_2$Te$_4$ (MBT) has been reported as an intrinsic magnetic topological insulator [24–32]. It is a layered compound with intralayer FM order and interlayer AFM order. A basic septuple-layer (SL) unit is formed by Te-Bi-Te-Mn-Te-Bi-Te consecutive atomic layers. Due to the van der Waals interaction between SL units, thin-layer samples can be obtained by the mechanical exfoliation method. The QAHI and Chern insulator states have been found in exfoliated MBT flakes with odd SLs [33–38], and high-Chern-number ($C = 2$) quantum Hall effect without Landau levels has also been reported in 9-SL MBT [34]. Additionally, the layer Hall effect and axion insulator state have been obtained in even-SL MBT devices [35,36]. Furthermore, it is predicted that CMEMs could emerge in the heterojunction of MBT and an *s*-wave superconductor [39]. All these facts manifest that MBT is an ideal platform for discovering exotic quantum phases. Compared with a magnetically doped topological insulator system, MBT has an ordered magnetic atom arrangement, larger magnetic energy gap, and less disorder, which may be a better system to host CMEMs. However, so far, there is little relevant experimental research on the coupling between MBT and superconductors.

Here, we fabricate the S-MBT-S junctions and perform magnetotransport experiments. By measuring the differential resistance ($dV/dI$) spectroscopy, it is found that proximity-induced superconductivity has been realized on the surface of MBT. A zero-bias $dV/dI$ peak appears with a pair of $dV/dI$ dips nearby, which indicate the proximity-induced superconducting gap $\Delta_{ind}$ in MBT. The induced gap increases gradually when the Fermi level approaches the charge neutral point (CNP). Moreover, $\Delta_{ind}$ exhibits an anomalous enhancement at finite magnetic fields, which may be related to the strong spin-orbit coupling (SOC) and magnetic structure of MBT.

[*]tanzb@sustech.edu.cn
[†]liaozm@pku.edu.cn

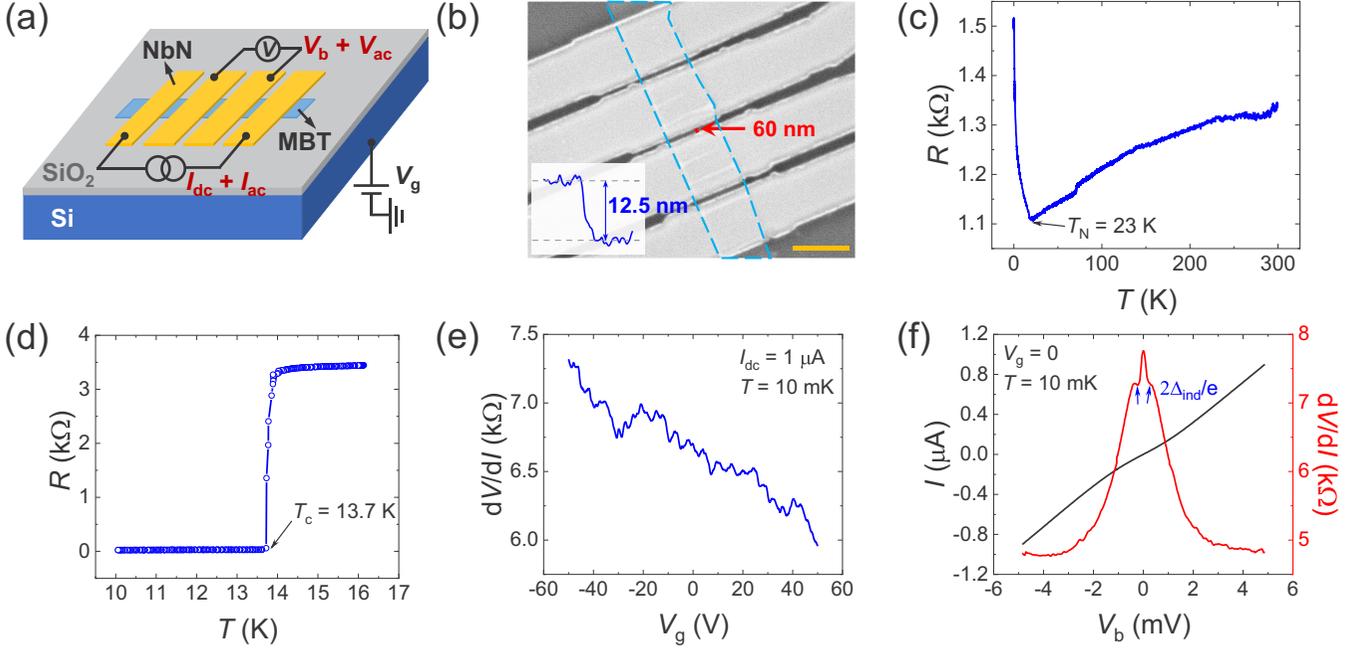

FIG. 1. Schematic and characterization of NbN-MBT-NbN junction. (a) Schematic of the NbN-MBT-NbN device and transport measurement configuration. The four-probe method is applied to measure the differential resistance spectroscopy. The $SiO_2$/Si substrate serves as a global back gate. (b) A scanning electron microscope (SEM) image of the device studied in this paper. The MBT thin flake is denoted by the cyan dashed line, and the width is ~1 μm. The channel length of the junction studied is ~60 nm (pointed at by the red arrow). Scale bar: 1 μm. Inset: Height profile of the MBT flake obtained by atomic force microscope. The thickness is measured to be ~12.5 nm (9 SLs). (c) The resistance-temperature (R-T) characteristic of a typical MBT sample. An inflection point at $T_N = 23$ K manifests the antiferromagnetic phase transition. (d) The R-T curve of the NbN electrode, showing a superconducting critical temperature $T_c = 13.7$ K. (e) Transfer curve of MBT at $T = 10$ mK with bias current $I_{dc} = 1$ μA. The MBT thin flake is n-type doped. (f) Current-voltage characteristic (black curve) and differential resistance spectroscopy (red curve). The nonlinear I-V curve shows a tunnelling behavior. The two differential resistance dips near zero bias indicate the proximity-induced superconducting gap $2\Delta_{ind}$ (denoted by the blue arrows).

## II. RESULTS AND DISCUSSION

A few-layer MBT flake was mechanically exfoliated and then transferred onto a silicon substrate with an oxide layer (~285 nm thick) serving as a back gate. The superconducting NbN electrodes were defined by standard electron beam lithography and grown by a dc magnetron sputtering technique, with niobium being the target material and a mixture of argon and nitrogen being the carrier and reaction gases, respectively. A schematic of the device is presented in Fig. 1(a). A four-probe method is applied throughout the transport study unless specified. A bias current $I_{dc}$ combined with a small ac excitation $I_{ac}$ (1 nA) was applied from source to drain. A standard lock-in amplifier (SR830) was used for detection of differential resistance. Figure 1(b) shows a scanning electron microscope image of the NbN-MBT-NbN junction, with the channel length ~60 nm. The thickness ~12.5 nm (9 SLs) of the MBT flake was measured with an atomic force microscope, as plotted by the inset of Fig. 1(b). The transport measurements were carried out in a commercial Bluefors dilution refrigerator with a base temperature of 10 mK equipped with a superconducting vector magnet.

Figures 1(c)–1(f) show the basic transport characteristics of the NbN-MBT-NbN heterojunction. With decreasing temperature, the resistance of MBT first declines at the high-temperature regime and exhibits a sharp increase below $T_N \sim 23$ K, which indicates the AFM transition of MBT at the Néel temperature [Fig. 1(c)]. The insulating behavior under $T_N$ may be ascribed to the magnetic impurities enhanced carrier scatterings [40,41]. The superconducting critical temperature $T_c$ of the NbN electrodes is ~13.7 K, as shown in Fig. 1(d), while the critical magnetic field of NbN is beyond the limit of the instrument (6 T) in this paper. The MBT sample is n-type doped, as shown by the transfer curve [Fig. 1(e)], in which the CNP is not reached even at $V_g = -50$ V, which may be due to the superconductor electrodes NbN induced electron doping and antisite defects naturally formed during the growth process [42]. At 10 mK, the current-voltage characteristic indicates a tunneling behavior [denoted by the black line in Fig. 1(f)], which is attributed to the potential barrier formed between the NbN electrodes and MBT thin flake due to charge transfer and band bending, considering their work function imbalance (NbN: 4.7 eV, MBT: 4.0 eV) [43,44]. Differential resistance is plotted by the red curve in Fig. 1(f). A pair of resistance dips symmetrically located around zero bias are observed (denoted by the two blue arrows), which represents the superconducting proximity-induced gap $2\Delta_{ind}$.

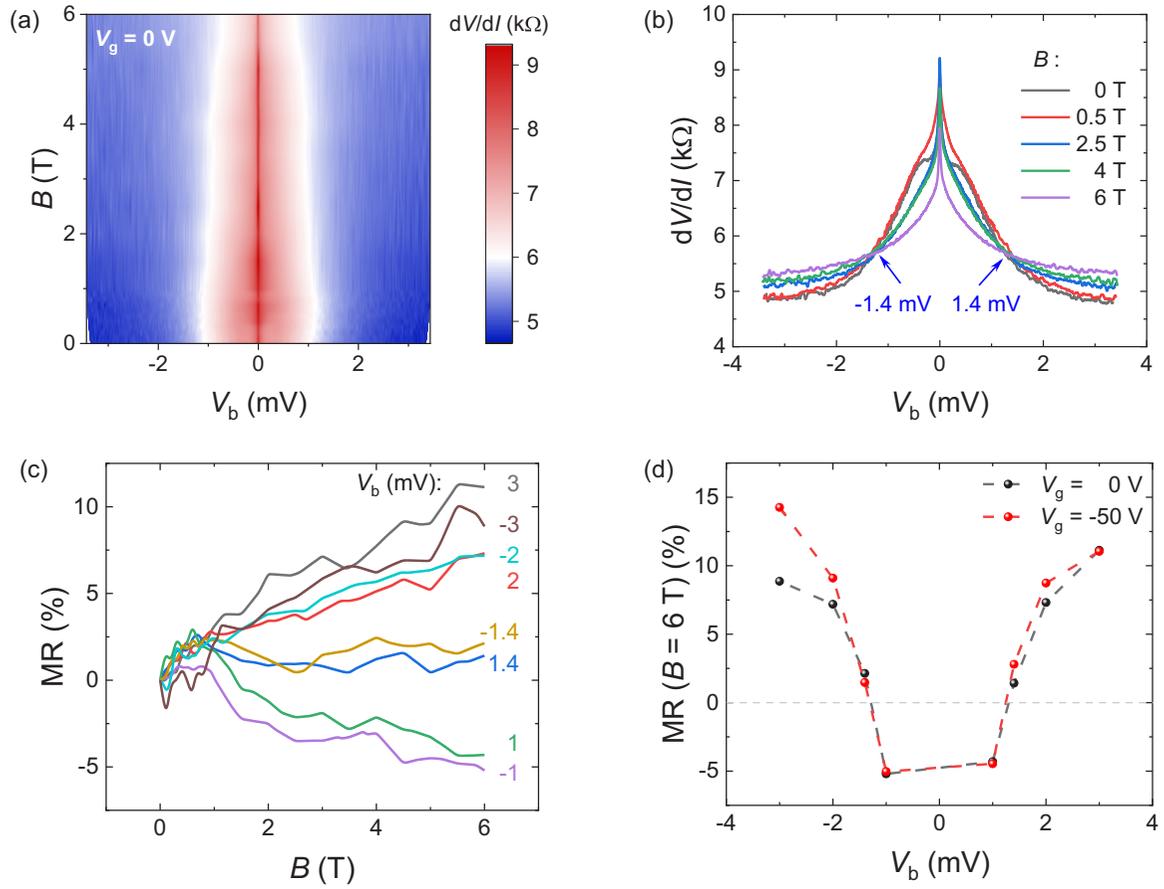

FIG. 2. Negative-to-positive magnetoresistance (MR) transition in NbN-MBT-NbN junction. (a) Two-dimensional (2D) color map of differential resistance as a function of magnetic field and bias voltage $V_b$ at $V_g = 0$ V. The central red region indicates the existence of potential barriers. (b) $dV/dI - V_b$ relation at various magnetic fields for $V_g = 0$ V. All curves cross approximately at bias voltage $V_b \sim \pm 1.4$ mV (denoted by the blue arrows), at which the MR is nearly zero. (c) MR of the heterojunction at different bias voltages and at $V_g = 0$ V, extracted from multiple vertical linecuts of Fig. 2(a). A negative-to-positive MR transition occurs when increasing the magnitude of bias voltage $V_b$. {MR = $[R(B) - R(0)]/R(0) \times 100\%$}. (d) Bias voltage dependence of MR at $B = 6$ T when $V_g = 0$ (black symbol) and $V_g = -50$ V (red symbol), respectively.

The $\Delta_{ind}$ is estimated to be ∼0.11 meV at 10 mK when $V_g = 0$ V.

Figure 2 shows the detailed magnetotransport properties of the junction. The magnetic field and bias voltage dependence of $dV/dI$ at $V_g = 0$ V is plotted in Fig. 2(a). The central vertical red region represents the resistance peak around zero bias, exhibiting the existence of potential barriers within the junction. Bias voltage dependence of $dV/dI$ under various magnetic fields ranging from 0 to 6 T is further displayed in Fig. 2(b). Lower resistance at greater bias voltage is observed as a result of general tunneling behavior at all magnetic fields. It is worth noting that different curves cross at $V_b \sim \pm 1.4$ mV, which indicates the transition from negative magnetoresistance (MR) to positive MR, as shown more apparently in Fig. 2(c). The MR at $B = 6$ T is about $-5\%$ and $10\%$ for $V_b = \pm 1$ and $\pm 3$ meV, respectively, where MR = $[R(B) - R(0)]/R(0) \times 100\%$. At the low-bias voltage regime, the reduced barrier height and enhanced tunneling process by the magnetic field may induce a negative MR [45]. On the other hand, when the applied bias exceeds the potential barrier, the intrinsic magnetotransport behavior of MBT may lead to a positive MR [31,46,47]. When tuning the back gate to $-50$ V, similar behavior shows up. The MR at $B = 6$ T is plotted as a function of bias voltage for $V_g = 0$ and $-50$ V in Fig. 2(d), respectively, exhibiting stable transition from negative MR to positive MR at different Fermi levels.

Gate control of the proximity-induced gap in the NbN-MBT-NbN junction is further studied, as displayed in Fig. 3. The bias dependence of differential resistance at various gate voltages is shown in Fig. 3(a). A pair of resistance dips symmetrically located near zero bias is obtained, which corresponds to the proximity-induced gap $2\Delta_{ind}$ in MBT surface beneath the NbN superconductors. The value of $\Delta_{ind}$ is determined by nodes of $dR/dV_b$. Gate evolution of the induced gap is denoted as dashed green lines, indicating the increase of $\Delta_{ind}$ when tuning the back gate to negative values, as shown in Fig. 3(b). The phenomenon could be explained as follows. When tuning the gate voltage to negative, the Fermi level approaches the CNP, and the original bulk-conduction-dominated regime gradually turns into a surface-state-dominated one. As the superconducting order parameter $\Delta$ decays rapidly in MBT bulk layers [schematically shown in Fig. 3(c)], it is not surprising that the induced gap becomes larger when gradually suppressing the bulk conduction at negative gate voltage, which is consistent with the experimental observations.

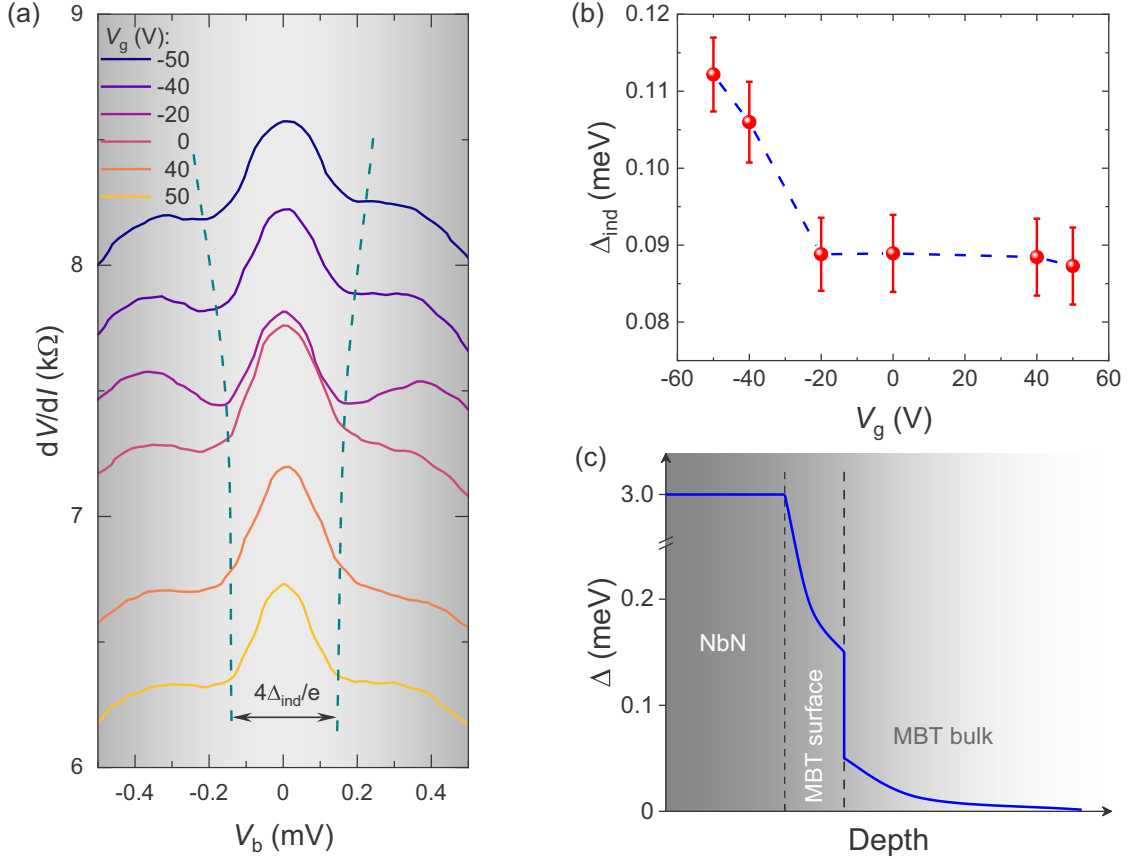

FIG. 3. Gate modulation of the proximity-induced superconducting gap when $B = 0$ T. (a) Differential resistance-bias voltage traces at various gate voltages. A pair of resistance dips symmetrically located around zero bias indicates the proximity-induced superconducting gap $2\Delta_{\text{ind}}$. The evolution of dip position with back gate is shown by the green dashed curves. (b) Gate voltage modulation of $\Delta_{\text{ind}}$. (c) Schematic of the spatial evolution of gap energy in the NbN-MBT heterostructure.

Furthermore, the magnetic field modulation of the induced gap $\Delta_{\text{ind}}$ is studied in Fig. 4. The measurements were carried out at $V_g = -50$ V to exclude the bulk conduction to the utmost. The CNP and p-type conduction are not reachable in this device due to the high electron doping from the electrodes and the short channel length of ∼60 nm. The bias and magnetic field dependence of $dR/dV_b$ is plotted as a color map in Fig. 4(a). The boundaries between the central red (blue) region and outer bright white regions represent the induced gap $2\Delta_{\text{ind}}$, at which resistance obtains a local minimum. Evolution of $dV/dI - V_b$ traces under multiple magnetic fields ranging from 0 to 0.9 T with equal spacing is plotted in Fig. 4(b) with offset for clarity. A pair of resistance dips is again obtained near zero bias, relating to the induced gap $2\Delta_{\text{ind}}$. The magnetic field evolution is denoted by the dashed green lines. The nonmonotonic behavior is further exhibited in Fig. 4(c), where $\Delta_{\text{ind}}$ reaches its maximum at a moderate magnetic field.

The anomalous enhancement of superconductivity at low magnetic field has been predicted and observed previously in other material systems [48–53], which is attributed to several possible origins: magnetic field modulation of the coherence length $\xi$ [48,52], strong SOC-induced spin-triplet component [52,53], field polarization of local magnetic moments [49–51], etc. To identify the origin of our results, we need to consider the options above. In clean samples with long mean free path, the field modulation of coherence length $\xi$ leads to enhanced superconductivity, according to the Kogan-Nakagawa mechanism [48]. However, it is easy to dismiss as the origin since our samples are not in the clean limit [heavy electron doping, Fig. 1(e)]. We do not exclude the probable influence of SOC since MBT does possess a strong SOC as an intrinsic magnetic topological insulator [47,54]. As a result, a spin-triplet part in $\Delta_{\text{ind}}$ may emerge due to the existence of SOC which tends to counter the pairing-splitting effect and stabilize superconductivity [52,55]. Indeed, enhancement of critical current $I_c$ [53] and induced gap $\Delta_{\text{ind}}$ [52] are reported in various systems possessing SOC. We thus consider SOC as one of the possible reasons in our experiments. Next, we analyze the impurity scattering. Considering the possible existence of magnetic impurity attached to the MBT surface during the fabrication process and the misalignment of the local magnetic moments, the magnetic field-enhanced superconductivity is also anticipated in MBT. Polarization of local magnetic moments under an external magnetic field reduces the exchange scattering rate of carriers, thus quenching the pair-breaking effect by a small magnetic field [49,51]. Formally, the exchange-scattering time is quantified by $\tau_B \propto E_F/2\pi J^2$, where $J$ is the exchange coupling and $E_F$ is the Fermi energy [51]. Therefore, the alleviation of the impurity scattering by a small external magnetic field can also result in the $\Delta_{ind}$ enhancement. When further increasing the magnetic field, the pair-breaking effect stands out as a detrimental factor to superconductivity, thus resulting in a decaying $\Delta_{ind}$ thereafter.

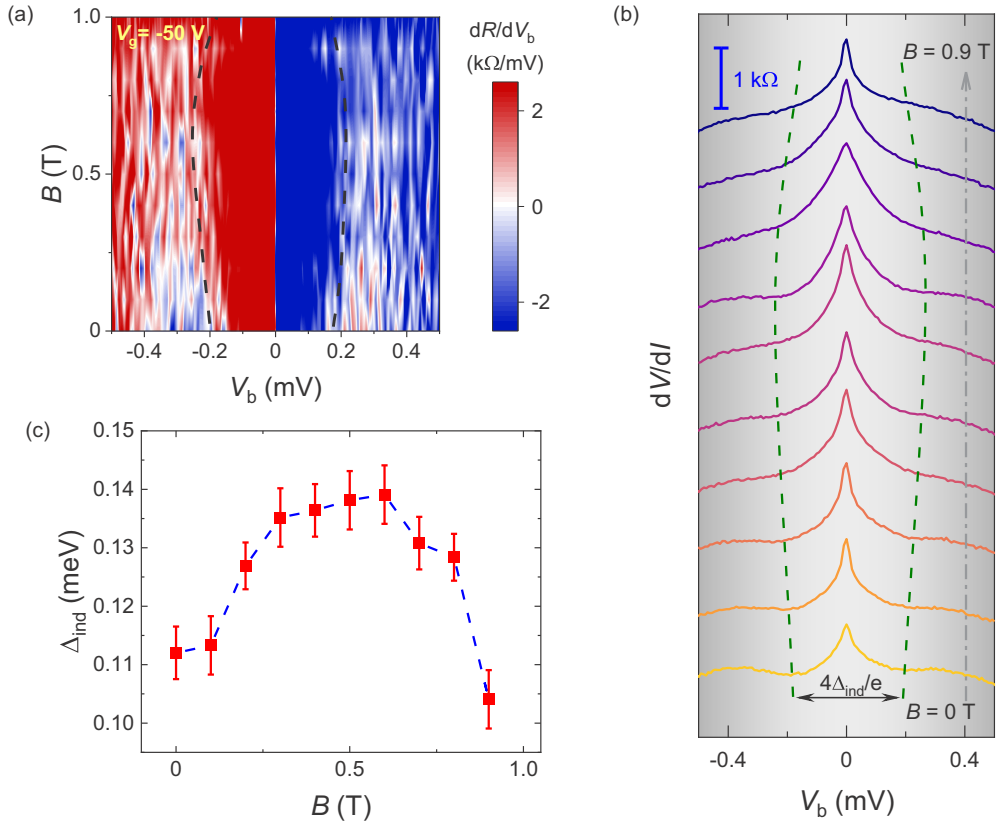

FIG. 4. Anomalous magnetic field-assisted enhancement of the proximity-induced superconducting gap at $V_g = -50$ V. (a) Derivative of resistance with respect to bias voltage ($dR/dV_b$) as a function of magnetic field and bias voltage. Boundaries between the central red (blue) region and the outer bright white area indicate the induced gap $2\Delta_{ind}$, as plotted by the black dashed curves. (b) $dV/dI - V_b$ traces at various magnetic fields ranging from 0 to 0.9 T with 0.1 T step. Traces are offset for clarity. The evolution of $\Delta_{ind}$ is displayed by the green dashed curves. (c) Magnetic field modulation of $\Delta_{ind}$.

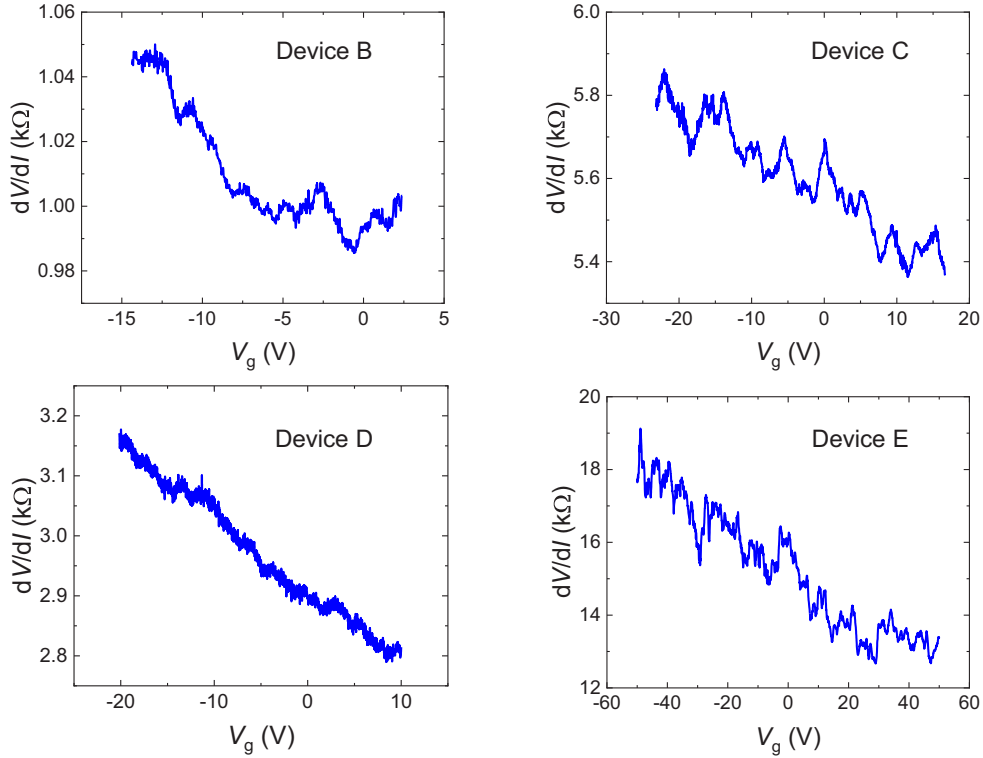

FIG. 5. Transfer curves for devices B–E at $T = 10$ mK. $n$-type doping is observed in all the samples.

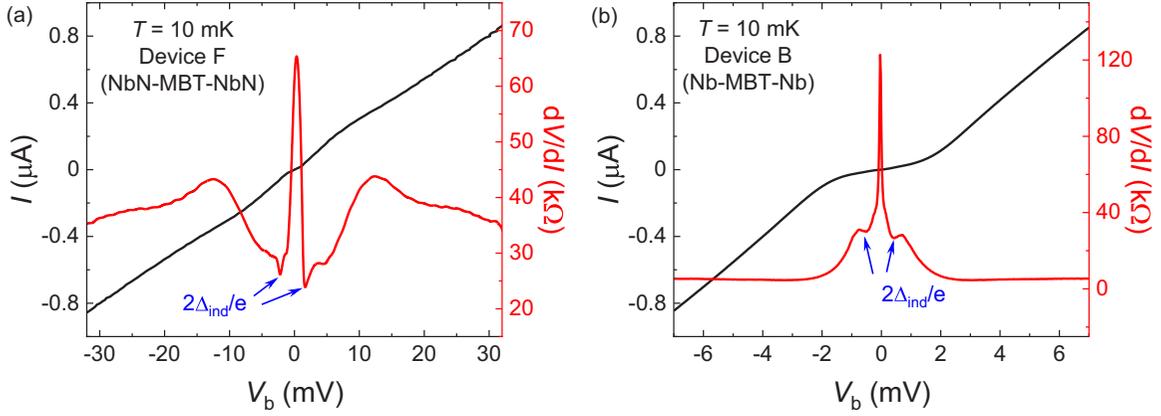

FIG. 6. Induced gap $\Delta_{ind}$ in (a) device F and (b) device B. $\Delta_{ind} \sim 0.96\,\text{meV}$ for device F and $0.23\,\text{meV}$ for device B.

### III. CONCLUSIONS

In summary, through magnetotransport measurements in the NbN-MBT-NbN junctions at low temperature, a proximity-induced superconducting gap $\Delta_{ind}$ is obtained, and its gate as well as magnetic field modulation is studied. The increase in $\Delta_{ind}$ when tuning the Fermi level toward the CNP is related to the enhanced interplay between superconductivity and electronic topology. Moreover, the anomalous enhancement of $\Delta_{ind}$ is observed at the small magnetic field regime, which is attributed to the strong SOC and polarization of the local magnetic moments of impurities attached to the sample. A transition from negative to positive MR is also achieved with increasing bias voltage applied across the junction. In this paper, we exhibit the interaction between magnetism and superconductivity in an intrinsic magnetic topological insulator MBT, which paves an avenue for further studies in heterostructures of QAHI and superconductors, seeking for the manipulation of CMEMs for the potential application in topological quantum computing.

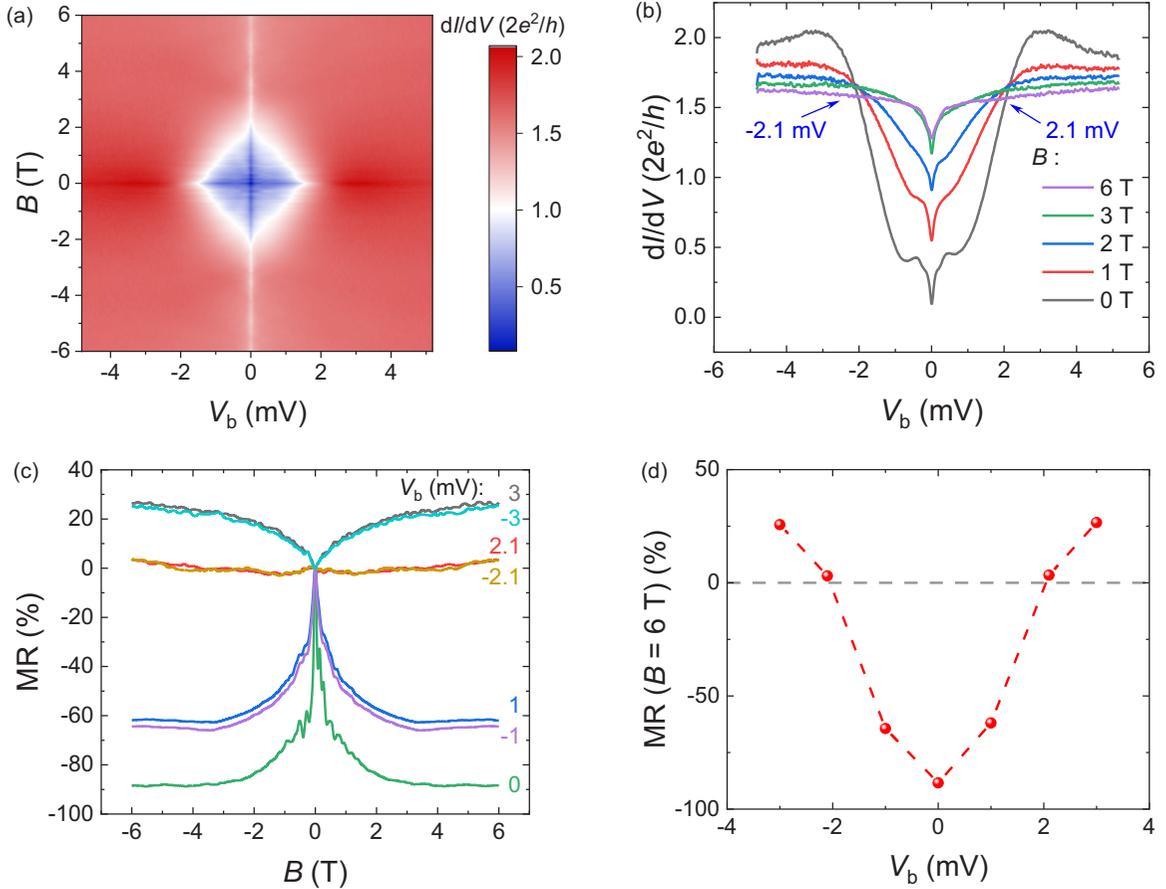

FIG. 7. Negative-to-positive magnetoresistance (MR) transition in device B (Nb-MBT-Nb). (a) Two-dimensional (2D) color map of differential conductance as a function of $B$ and $V_b$. (b) $dI/dV - V_b$ curves at different magnetic fields for $V_g = 0\,\text{V}$. All curves cross approximately at bias voltage $V_b \sim \pm 2.1\,\text{mV}$ (denoted by the blue arrows). (c) MR of device B at different $V_b$, extracted from multiple vertical linecuts of Fig. 7(a). A negative-to-positive MR transition occurs when increasing the magnitude of bias voltage $V_b$. {MR = $[R(B) - R(0)]/R(0) \times 100\%$}. (d) $V_b$ dependence of MR at $B = 6\,\text{T}$.

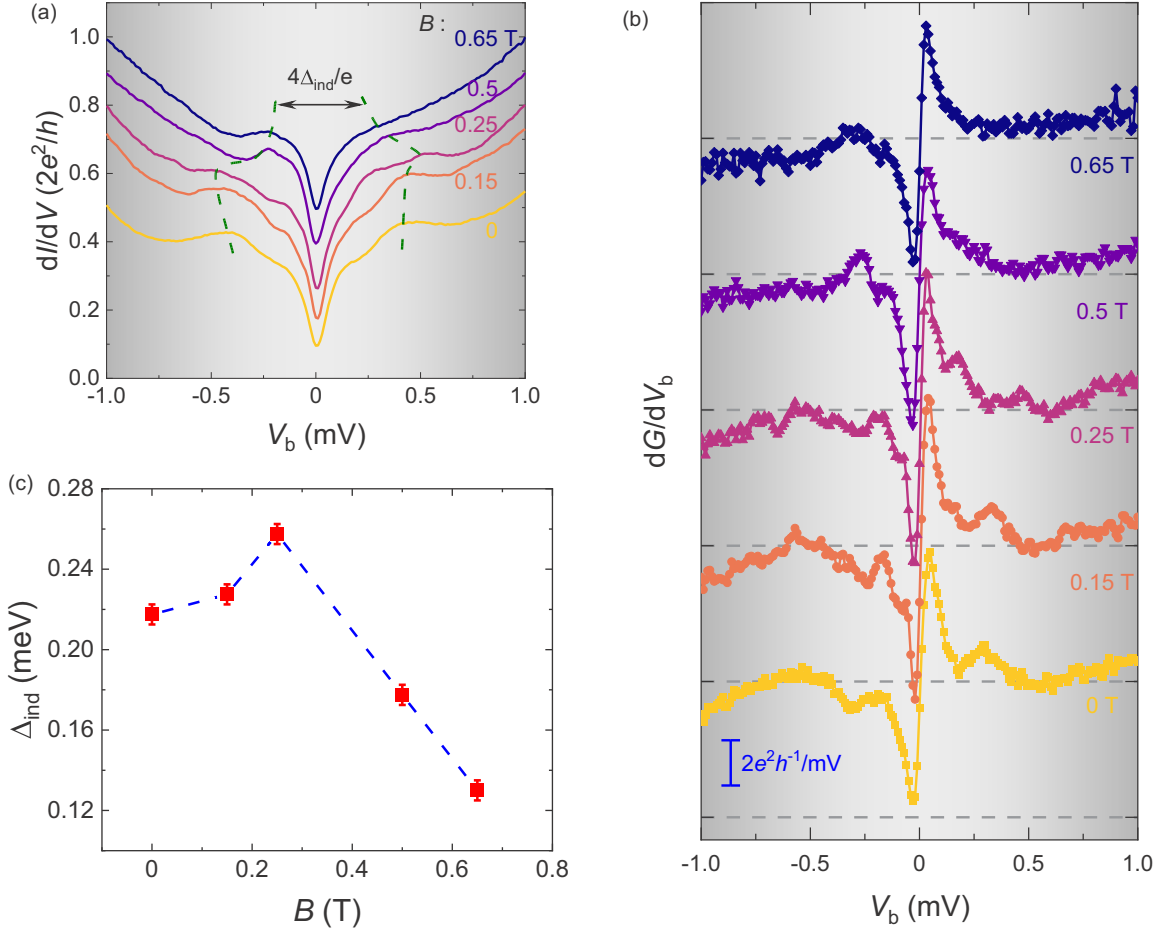

FIG. 8. Anomalous magnetic field enhancement of $\Delta_{ind}$ in device B. (a) $dI/dV - V_b$ traces at various magnetic fields ranging from 0 to 0.65 T. The evolution of $\Delta_{ind}$ is shown by the green dashed curves. (b) $dG/dV_b$ traces at corresponding magnetic fields. Curves are shifted for clarity. (c) Magnetic field modulation of $\Delta_{ind}$. An anomalous enhancement is obtained at low magnetic field.

## ACKNOWLEDGMENTS

This paper was supported by the Key-Area Research and Development Program of Guangdong Province (Grants No. 2020B0303060001 and No. 2018B030327001), the National Key Research and Development Program of China (Grant No. 2018YFA0703703), the National Natural Science Foundation of China (Grants No. 91964201 and No. 61825401), and the Guangdong Provincial Key Laboratory (Grant No. 2019B121203002).

## APPENDIX A: $n$-TYPE DOPING BEHAVIOR IN DEVICES B–E

Efforts have been made to detect the proximity-induced superconducting gap $\Delta_{ind}$ in the $p$-type region of MBT. In our previous devices consisting of normal metal electrodes (e.g., Ti/Au) grown by electron beam evaporation with channel length $\sim 1$ $\mu$m, the $p$-type conduction was achieved [40]. However, due to the heavy electron doping effect of Nb or NbN superconducting electrodes grown by magnetron sputtering [56], as well as the relatively short channel length ($\sim 60$ nm), devices B–E (devices B–D: Nb-MBT-Nb; device E: NbN-MBT-NbN) show a similar $n$-type conduction as the device studied in the main text, as depicted in Fig. 5.

## APPENDIX B: REPRODUCIBLE PROXIMITY-INDUCED GAP $\Delta_{ind}$ IN MBT JUNCTIONS

$I$-$V$ characteristics and differential resistance curves in device F (NbN-MBT-NbN) and device B (Nb-MBT-Nb) are depicted in Fig. 6. Repeatable $dV/dI$ dips symmetrically located around zero bias are observed in device F, indicating the induced gap $\Delta_{ind} \sim 0.96$ meV. Tunneling behavior is obtained in device B, and it again possesses resistance dips corresponding to $\Delta_{ind} \sim 0.23$ meV, manifesting that the superconducting proximity effect is also achieved by the Nb electrodes.

## APPENDIX C: NEGATIVE-TO-POSITIVE MR IN Nb-MBT-Nb JUNCTION

Figure 7 exhibits the MR transition in device B. Here, $dI/dV$ as a function of $B$ and $V_b$ is plotted as a two-dimensional color map in Fig. 7(a). Extracting several horizontal linecuts, $dI/dV$ at different $B$ fields is obtained [Fig. 7(b)]. Here, $dI/dV$ peaks are clearly located symmetrically around zero bias when $B = 0$ T, indicating the induced gap $2\Delta_{ind}$. Peaks disappear at higher magnetic fields, and the traces approximately cross at $V_b = \pm 2.1$ mV. MR at different $V_b$ is calculated [Fig. 7(c)], and the negative-to-positive MR transition is clearly seen [Fig. 7(d)].

## APPENDIX D: ANOMALOUS MAGNETIC-FIELD-ENHANCED $\Delta_{ind}$ IN Nb-MBT-Nb JUNCTION

The evolution of $\Delta_{ind}$ under the magnetic field in device B is studied (Fig. 8). Differential conductance peaks are obtained at low magnetic field [Fig. 8(a)], and the derivative $dG/dV_b$ is calculated to determine the gap value [Fig. 8(b)]. Indeed, consistent with the NbN-MBT-NbN junction studied, an anomalous enhancement in $\Delta_{ind}$ is observed, and the maximum occurs at $\sim 0.25$ T [Fig. 8(c)].